\documentclass[conference]{IEEEtran}
\usepackage[all]{nowidow}
\usepackage{balance}
\ifCLASSINFOpdf
   \usepackage[pdftex]{graphicx}
   \graphicspath{{../pdf/}{../jpeg/}}
   \DeclareGraphicsExtensions{.pdf,.jpeg,.png}
\else
   \usepackage[dvips]{graphicx}
   \graphicspath{{../eps/}}
   \DeclareGraphicsExtensions{.eps}
\fi
\begin{document}
%
\title{Feature Extraction in Augmented Reality}

\author{\IEEEauthorblockN{Jekishan K. Parmar}
\IEEEauthorblockA{Department of Computer Science \& Engineering,\\
Babaria Institute of Technology,\\
Vadodara\\
Email: jekishan@aol.in}
\and
\IEEEauthorblockN{Ankit Desai}
\IEEEauthorblockA{IQM Corporation \&\\
School of Engineering \& Applied Science,\\
Ahmedabad University, Ahmedabad\\
Email: desaiankitb@gmail.com}}


%


\maketitle

\begin{abstract}
Augmented Reality (AR) is used for various applications associated with the real world. In this paper, first, describe characteristics and essential services of AR. Brief history on Virtual Reality (VR) and AR is also mentioned in the introductory section. Then, AR Technologies along with its workflow is depicted, which includes the complete AR Process consisting of the stages of Image Acquisition, Feature Extraction, Feature Matching, Geometric Verification, and Associated Information Retrieval. Feature extraction is the essence of AR hence its details are furnished in the paper.
\end{abstract}


%
\IEEEpeerreviewmaketitle

\section{Introduction}
\begin{figure}[!h]
\centering
\includegraphics[scale=0.35]{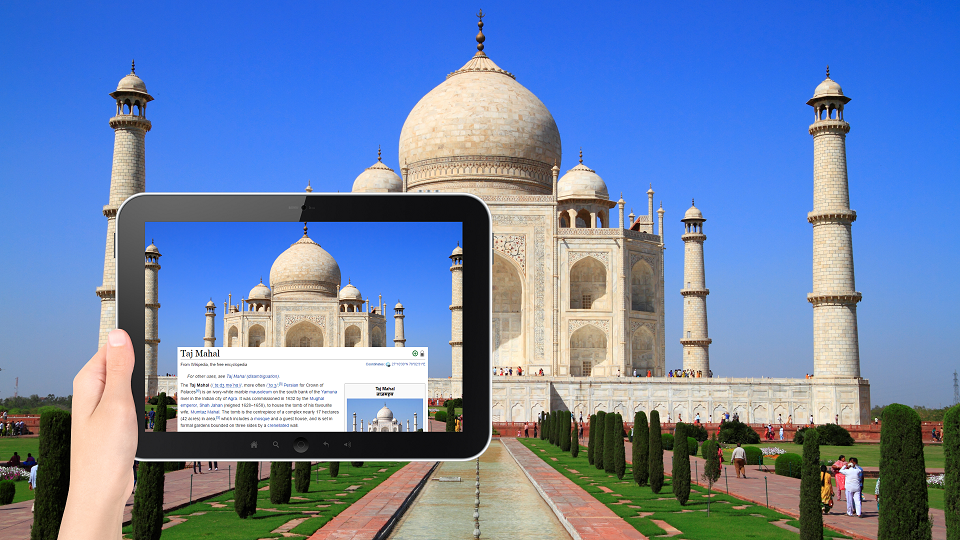}
\caption{Augmented Reality camera example\protect\footnotemark}
\label{fig1:Augmented Reality camera example}
\end{figure}

The term Augmented Reality holds multiple definitions but one of the most popular and the one that still holds is Augmented reality is a field in which 3D virtual objects are integrated into a 3-D real environment in real time \cite{paper1}. Though, AR can be looked from the point of view of enhancement to reality, it can be defined as AR is use of computers to enhance the richness of the real world \cite{paper2}. It can also be simply called user interaction in 3D environment \cite{paper3}. When it is perceived as an interaction between humans and virtual objects, it can also define it as combining the  real and the virtual in order to assist the user in performing a task in a physical setting is called Augmented Reality \cite{paper4}. According to \cite{paper5}, Augmented Reality supports context-aware computing. For an AR user, the real world and virtual objects coexist on the same view. For example, in a museum there is Mona Lisa, when this picture of Mona Lisa is looked through an augmented reality smartphone, the information of Mona Lisa will be imposed on top. In addition, it may include information like the artist, da Vinci, as well as the year that it was estimated that this picture was drawn. Figure 1, shows one such example of augmented reality camera with the Taj Mahal. It can be seen in the figure \ref{fig1:Augmented Reality camera example} that as the camera is acquiring the Taj in its lenses and displaying to its user, information related to the Taj is also being displayed to the user.

Next section compares AR and VR with deep insights into current advancements and research challenges, section III depicts various technologies involved in AR, then, details about Feature Extraction is furnished in section IV and finally, section V concludes the paper. 
 

\section{Augmented Reality And Virtual Reality}
Jaron Lanier coined the phrase “Virtual Reality”, in 1989 \cite{paper6} and Thomas Caudell defined “Augmented Reality” in 1990 \cite{paper7}. Hence, these concepts are quite old. Though, its complete implementation has become possible only recently and that too, with full reliability.

\subsection{Reasons for Recent advancements in Augmented Reality}

\begin{itemize}
\item High resolution cameras, which enables accurate image and object identification, are only available since recent times. They are also available in smart devices.
\item Recent availability of very high performing Central Processing Units and Graphical Process Units due to this fast and reliable image processing along with feature extraction can be done, which is necessary for Augmented Reality.
\item Cheap availability of large amount of memory and faster input/output access which is useful to store object information and quickly access it, which is highly essential for AR.
\item Sharp virtual text, and sharp images, superimposed on smart devices in a very elegant, and yet, easy-on-the-eye fashion has become possible only due to availability of High Definition displays on these devices.
\item Information needs to be retrieved speedily and brought to the device from AR servers and databases so that it can be displayed on this device for the user to be able to see quickly, which is recently possible due to advent of high-speed broadband, wireless and wired networking technologies.
\end{itemize}
Due to all these advancements, AR is going to become better and quicker. In spite of all these technological advancements there are some challenges in AR that needs to be worked on and requires proper study as well as research in order to get more from AR.

\subsection{Research Challenges in Augmented Reality}
Despite the growing interests and development in AR, following are the challenges that exists in the field which has very wide scope of research.
\begin{itemize}
\item A precise, fast, and robust registration between synthetic augmentations and the real environment is one of the most important challenge \cite{paper8}.
\item AR system needs to deal with vast amount of information that exists in the real world which requires very quick and portable hardware \cite{paper9}.
\item Highly efficient energy consumption from AR devices is desired due to their limited battery life \cite{paper9}.
\item High network dependency is also a challenge as there may not be network coverage in some areas \cite{paper10}.
\item AR assumes everything to be static while real world is very dynamic, as there are many things that keep on changing like people passing by, weather conditions, color of buildings as they may be repainted over a gap of few years. Thus, inability of AR to lack dynamism is also a challenge \cite{paper10}.
\end{itemize}

\footnotetext{http://www.arlab.com/img/content/products/matching01.jpg}
\subsection{Comparison of Augmented Reality and Virtual Reality}
A virtual reality user will be fully immersed in animated environment. This is like the game playing space that is commonly used. A user or a game player will commonly use an avatar to exist and interact inside the virtual space. The view of the user in virtual reality is different from the real environment and fantasies and illusions surrounding the avatar are easy to create in the virtual world.  Whereas, augmented reality is a mixture of real life and virtual reality \cite{paper11}.  This is how, it differs from virtual reality. So, augmented reality combines information and does various things, but its basis is real life, what the user actually sees \cite{paper12}.

\begin{figure}
\centering
\includegraphics[scale=0.48]{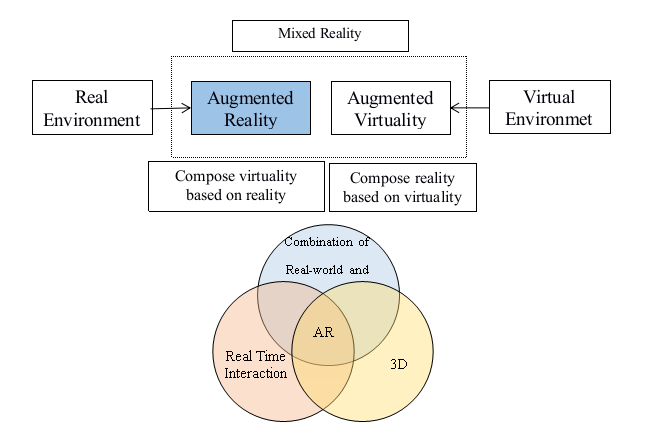}
\caption{Augmented Reality (AR) versus Virtual Reality (VR)}
\label{fig2:arvsvr}
\end{figure}

An augmented reality user is able to obtain useful information about location or objects and can interact with virtual contents in the real world. Therefore the virtual contents are superimposed onto the real world image that is seen by the user. The augmented reality user can distinguish the superimposed virtual objects and hence able to turn on or turn off selected AR functions, which may be related to certain objects. In other words, if a user is only interested in objects, and not in locations. In that case, user will go and adjust  augmented reality functions such that location information does not pop up, on their user screen and only gets object information. Things like that are controllable in futuristic systems. In comparison to virtual reality, augmented reality users commonly feel less separated from the real world, because basically the foundation of their view is exactly the real world. Things are just superimposed on the real world view of that user. So fantasies and illusions can be created and superimposed on the real world view.

Some of the definitions of AR has already been described, but its definition in context of virtual reality can be formed and compared with the definition of virtual reality itself. The comparison is shown in figure \ref{fig2:arvsvr}.

It can be seen in figure \ref{fig2:arvsvr}, that on one side there is the real environment and on the other side is the virtual environment, then these two combine into a mixed reality, where the augmented reality is virtually based on reality that is based on virtuality and augmented virtuality is based on virtuality which in turn is based on reality. Hence, augmented reality is based on reality where virtual information and images are overlapped and superimposed in the right position.

\section{Augmented Reality: Technology}
Figure 3 shows technological components present in augmented reality. On the left, there is content provider server which contains contents viz. 3D geographical assets, geographical information, text, images, movies and point of interest information. On the other side there are three major units to make augmented reality possible and this includes detection and tracking engines, a rendering system, and also interaction devices.

Detection and Tracking uses computer vision, tracking sensors, camera API and markers and features in order to carry out its task of Recognition. Necessary visualization is provided to the user by using computer graphics API, video composition and depth composition in order to render images.

\begin{figure}
\centering
\includegraphics[scale=0.5]{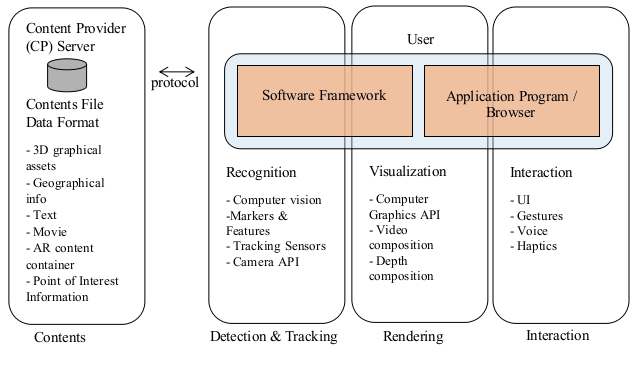}
\caption{Technological Components in Augmented Reality}
\label{fig3:Technological Components in Augmented Reality}
\end{figure}

Then, interaction is provided to the user by picking up gestures, voice, haptics and providing a UI. Thus, these three are the combination of software framework and application programming, including browser access, as shown in figure \ref{fig3:Technological Components in Augmented Reality}.

\subsection{Workflow of AR}
Fig 4 shows, the user sends an input, and it is detected by the interaction unit, then the interaction unit sends information to the rendering device. Now, the rendering device will take combined input from the detection and tracking unit. In addition, the contents needs to be streamed together and this, combined together by the rendering unit which will send back visual feedback to the user.

\begin{figure}
\centering
\includegraphics[scale=0.45]{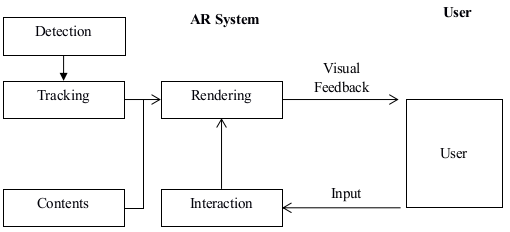}
\caption{AR Workflow}
\label{fig4:AR Workflow}
\end{figure}

\begin{figure}
\centering
\includegraphics[scale=0.09]{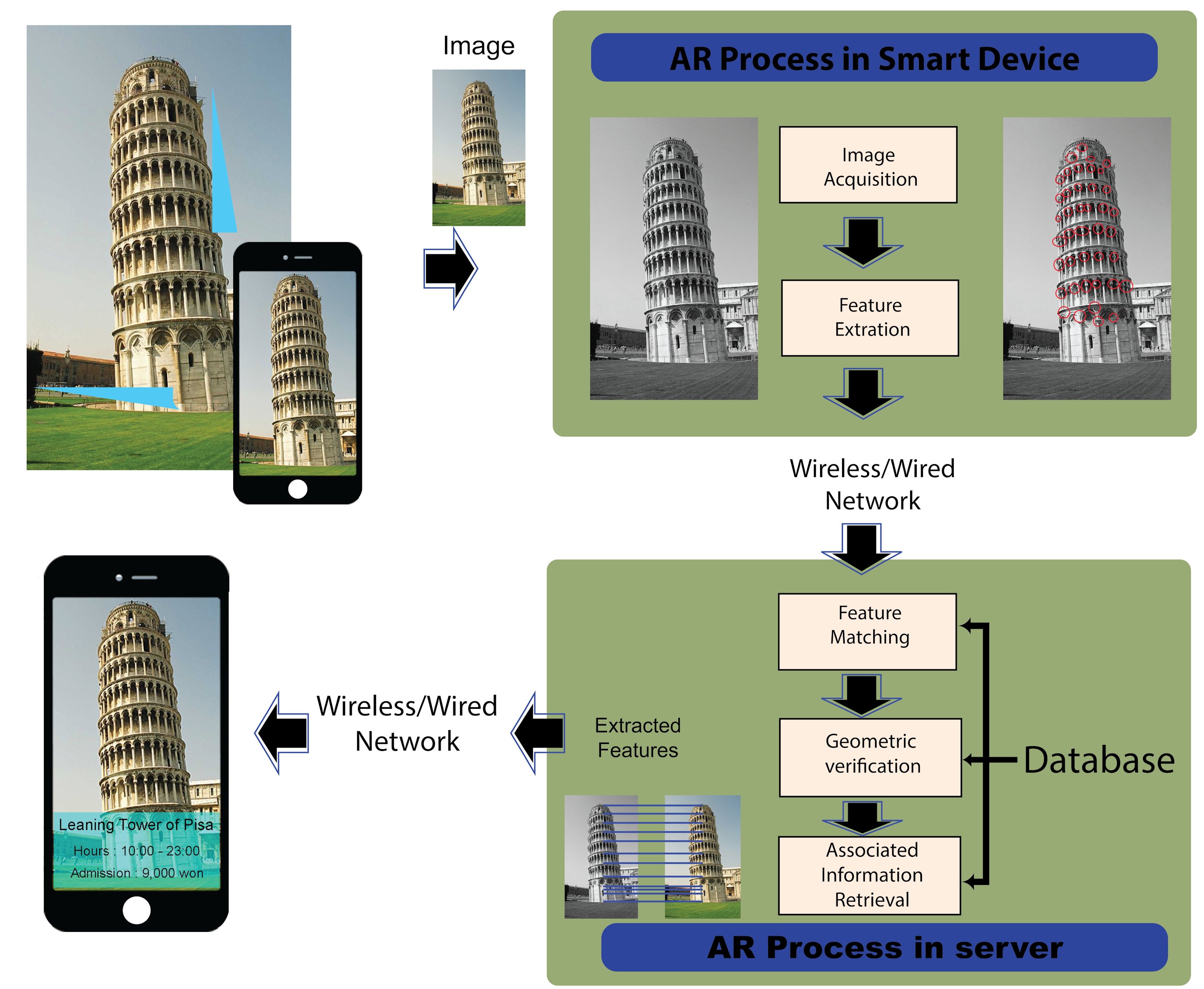}
\caption{Augmented Reality Process Example}
\label{fig5:Augmented Reality Process Example}
\end{figure}

\subsection{AR Process}
AR Process is divided into five steps, which can be described by an example.
\begin{itemize}
\item Image Acquisition: In this step, the image is retrieved from the augmented reality camera. In our example i.e. the picture which is shown in figure \ref{fig5:Augmented Reality Process Example}, is at the image acquisition stage and it is the Leaning Tower of Pisa in the middle of the device display.
\item Feature Extraction: It is based on an initial set of measured data. The extraction process generates informative non redundant values to cilitate the subsequent feature learning and generalization steps. As depicted in figure \ref{fig5:Augmented Reality Process Example}, part 2), there are redundant red circles on top of the image, which denotes extraction of features. Once the feature is extracted, these feature extracted points will be used to identify and learn what this object is.
\item Feature Matching: This is a process of computing abstractions of image information and to make a local decision if there is an image feature or not and this is conducted for all image points. In order to identify extracted points, other features are needed to be matched, which is performed in this step and the matched objects are sent for verification in next step.
\item Geometric Verification: This is an identification process of finding geometrically related images in the image data set. The image data set is a subset of the overall augmented reality image database. Matched objects in previous step are verified in this step by comparing it with objects in the database and are further sent for retrieval in next step.
\item Associated Information Retrieval: This process is to search and retrieve metadata, text, and content-based indexing information of the identified image or object. Associated information is used for display on the augmented reality screen near the corresponding image or object. As shown in the figure \ref{fig5:Augmented Reality Process Example}, in the middle lower part, there is Leaning Tower of Pisa and its information is superimposed on top of the image that is used on the Smartphone.
\end{itemize}

\begin{figure*}
\centering
\includegraphics[scale=0.26]{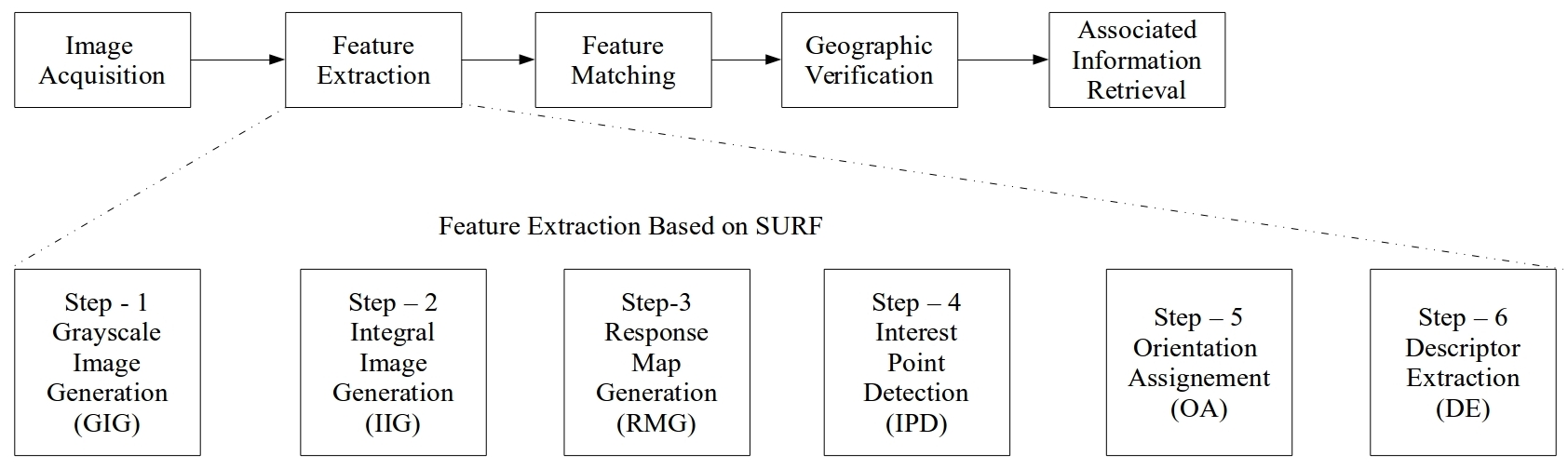}
\caption{Augmented Reality Process Example}
\label{fig6:Augmented Reality Process Example}
\end{figure*}

\section{Feature Extraction Process In Augmented Reality}
Feature Extraction is the second step in AR Process, though it composes of six stages as shown in figure \ref{fig6:Augmented Reality Process Example}. This section will discuss it in detail.
\begin{itemize}
\item Grayscale Image Generation (GIG): Here, the original image captured by the augmented reality device is changed into a grayscale value image in order to make it robust to colour modifications. 
\item Integral Image Generation (IIG): This is a process of building an integral image from the grayscale image. This procedure enables fast calculation of summations over image sub-regions.
\item Response Map Generation (RMG): In order to detect Interest Points (IPs)using the determinant of the Hessian Matrix of image, the RMG Process constructs the scale-space of the image. Only after having IPs, various operations can be done which leads close to actual feature extraction.
\item Interest Point Detection (IPD): Based on the generated scaled response maps, the maxima and minima are detected and used as the Interest Points.
\item Orientation Assignment (OA): Here, each detected Interest Point, is assigned a reproducible orientation to provide rotation invariance. Rotation invariance means invariance to image rotation.
\item Descriptor Extraction (DE): This is a process of uniquely identifying an interest point such that it is distinguished from other Interest Points.
\end{itemize}

\subsection{Feature Extraction with an Example}

\begin{figure}[h]
\centering
\includegraphics[scale=0.8]{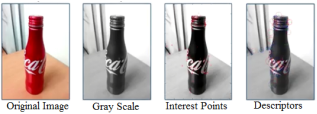}
\caption{Feature Extraction Example}
\label{fig7:Feature Extraction Example}
\end{figure}

The feature extraction process is about finding the Interest Points from the image or the video, detecting the descriptors, from the interest point and then compare the descriptors with the data in the database. As shown in figure \ref{fig7:Feature Extraction Example}. Here, first is the original image, then gray scale image, which is processed into the interest points, and as shown, there are certain locations, where some specific characteristics are identified and marked. Then in the next processing stage and based on some color coordination, the descriptors are found. With these descriptors  database can be queried and find matching information of what this object is. As shown here in figure \ref{fig7:Feature Extraction Example}, in order to have accurate descriptors, keen and sharp image is needed. Therefore, now with new enhanced cameras on smart devices, more accurate descriptors are found and therefore, the augmented reality information is more reliable, and there are less errors, and also it is processed much more quickly.

\subsection{Blob Detection}
Feature extraction is a combination where qualification for descriptors is needed in a very accurate way. Therefore, invariability from noise, scale, and rotation needs to be kept. In addition, there are various kinds of descriptors, e.g. corner descriptors, blob descriptors and region descriptors. In this section, describes blob detection with the help of an example. 

\begin{figure}[h]
\centering
\includegraphics[scale=0.2]{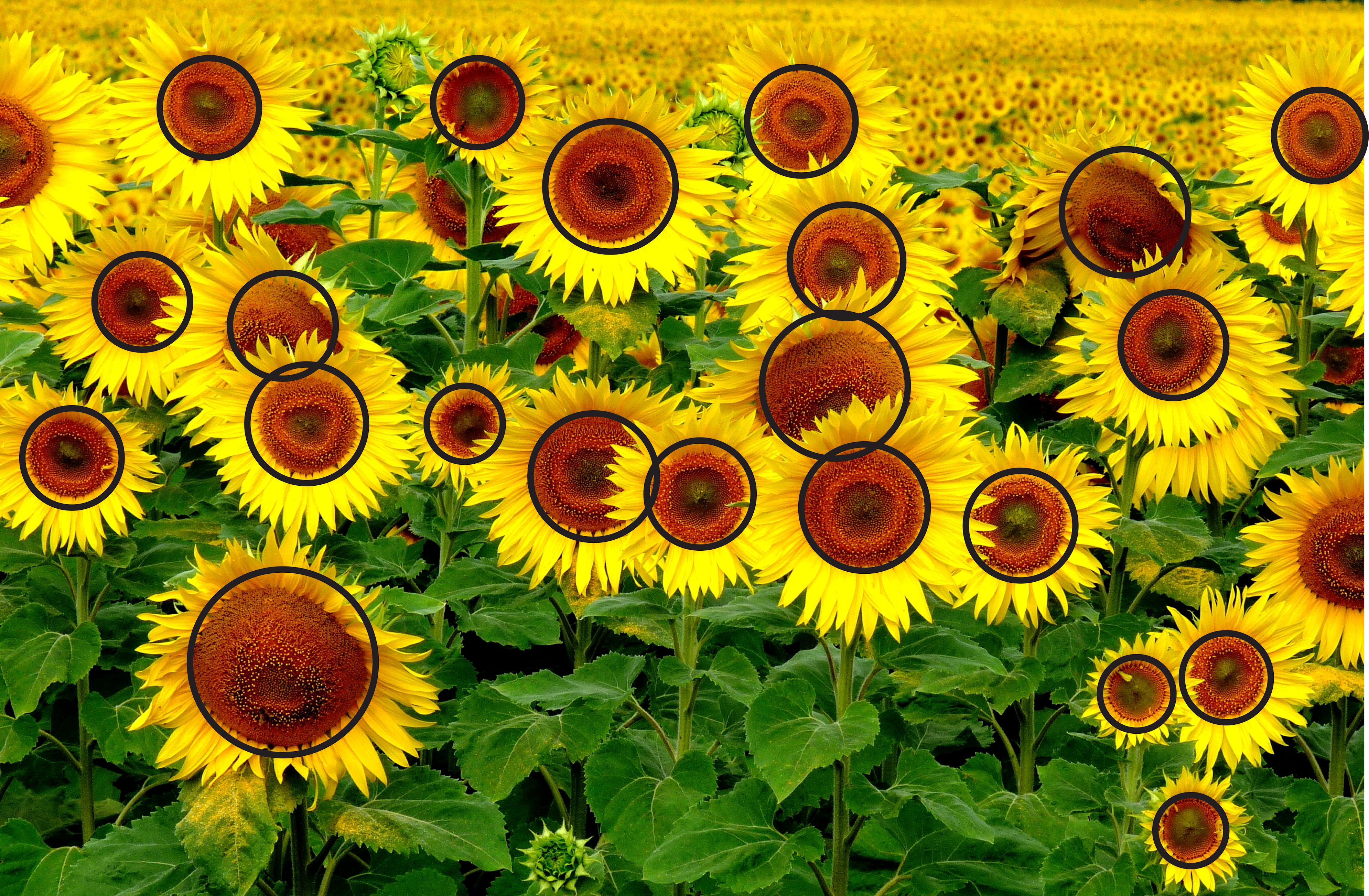}
\caption{Blob Detection Example}
\label{fig9:Blob Detection Example}
\end{figure}
A blob is a region of an image that has constant or approximately constant image properties. All the points in a blob are considered to be similar to each other. These image properties, which include brightness and color, are used in the comparison process to surrounding regions \cite{paper13}.

\subsection{Algorithm for Blob Detection}
Following are the steps are needed to be follow for Blob Detection \cite{paper14}. Figure 9, shows an image with its blobs detected.
\begin{itemize}
\item Start from the first line of the image and find groups of one or more white (or black) pixels, known as lineblobs.
\item Find X, Y co-ordinates of each those blob and number each of these groups.
\item Repeat this sequence on next line.
\item While collecting the lineblobs, check whether all lineblobs are collected and whether they overlap or not.
\item If lineblobs are overlapping then merge these lineblobs by using there X and Y co-ordinates to one blob and treat it as a whole blob.
Repeat this for every line and you have a collection of blobs.
\end{itemize}

\subsection{Other Feature Extraction Techniques}
Following are some feature extraction techniques:
\begin{itemize}
\item Haar \cite{paper15}
\item Scalable Invariant Feature Transform (SIFT) \cite{paper16}
\item Histogram of Oriented Gradient (HOG) \cite{paper17}
\item Speeded Up Robust Features (SURF) \cite{paper18}
\item Oriented FAST and rotated BRIEF (ORB) \cite{paper19}
\end{itemize}

SIFT and SURF will be described in detail in this section. SIFT is the most widely used feature extraction algorithm. It extracts features from images accurately and efficiently. It overcomes the various adverse effects of extraction, such as transformation, noise, and lightness. Following is the four step SIFT algorithm.

\begin{itemize}
\item Step 1. Scale space extreme detection
\item Step 2. Key point localization and filtering, 
\item Step 3. Orientation assignment, 
\item Step 4. Descriptor construction
\end{itemize}

SURF is an improvement over SIFT from the aspect of speed. Its algorithm is based on the same algorithmic principle as SIFT, but uses procedures that require less computation to enhance the processing speed. Therefore, SURF made it possible to carry out feature extraction in a near real-time or a real-time manner. Following is the three step algorithm of SURF:
\begin{itemize}
\item Step 1. Interest Point Detection, which is about high-speed detection of interest points.
\item Step 2. Local Neighbourhood Description, which is about descriptors using response of the Haar-wavelet.
\item Step 3. Matching Process, where faster matching algorithm is applied by using a Laplacian operator.
\end{itemize}

\section{Conclusion}
Looking at the recent advancements in AR Technologies viz. accurate cameras and smart devices, it is bound that more and more AR based devices and applications are going to capture its market share. As a pillar to that, the core component of AR i.e. Feature Extraction is going to be technologically challenging. For the same, SIFT and SURF, presented in this paper, are ready to be used as feature extraction algorithms. Apart from that, some more insight can be given into other feature extraction techniques that are described in the previous section. Though, emphasis on accuracy of the feature extraction and matching process is still a field of research which can be worked on by using new machine learning techniques and algorithms.






%
\balance

\balance

\end{document}